\documentclass[aps,showpacs,pra,twocolumn,superscriptaddress,floatfix,longbibliography]{revtex4-1}

\usepackage{amsthm}
\usepackage[utf8]{inputenc}
\usepackage{amsmath,amssymb,amsmath,amsthm,bm,graphicx,xcolor}
\usepackage{natbib}
\usepackage[breaklinks]{hyperref}
\usepackage{color}
\usepackage{tabularx}
\usepackage{tikz}
\usepackage{epsfig}
\usepackage{amssymb}
\usepackage{graphicx}
\usepackage{wasysym}
\usepackage{dcolumn}
\usepackage{epstopdf}
\usepackage{subfigure}
\usepackage{psfrag}
\usepackage{bbold}
\usepackage[breaklinks]{hyperref}
\hypersetup{colorlinks=true}

\newcommand{\up}{\uparrow}
\newcommand{\dn}{\downarrow}
\newcommand{\vecform}{\bm} 
\newcommand{\xx}{\vecform{x}}

\newcommand{\bra}[1]{\mbox{$\langle #1 |$}}

\newcommand{\ket}[1]{\mbox{$| #1 \rangle$}}

\newcommand{\E}{\mathcal{E}}        %% functional
\newcommand{\rr}{\bm{r}}         
   
\renewcommand{\H}{\mathcal{H}}
\begin{document}

\title{Static correlated 
functionals for reduced 
density matrix functional 
theory}
  
\author{Carlos L. Benavides-Riveros}
\email{carlos.benavides-riveros@physik.uni-halle.de}
\affiliation{Institut f\"ur Physik, Martin-Luther-Universit\"at
Halle-Wittenberg, 06120 Halle (Saale), Germany}

\author{Miguel A. L. Marques}
\affiliation{Institut f\"ur Physik, Martin-Luther-Universit\"at
Halle-Wittenberg, 06120 Halle (Saale), Germany}

\date{\today}

\begin{abstract}
Based on recent progress on fermionic exchange symmetry we 
propose a way to develop new functionals for reduced
density matrix functional theory. For some settings with an odd 
number of electrons, by assuming saturation of the inequalities stemming from the generalized Pauli principle, the many-body 
wave-function can be written explicitly in terms of the natural 
occupation numbers and natural orbitals. 
This leads to an expression for the 
two-particle density matrix and therefore for the 
correlation energy functional. This functional was then
tested for a three-electron Hubbard model where it showed 
excellent performance both in the weak and strong 
correlation regimes.
\end{abstract}

\maketitle

\section{Introduction}

The quantum many-body problem (the problem of computing 
the ground-state features of a system of many interacting
electrons) is at the very heart of quantum chemistry and condensed
matter physics. The complexity of such a problem is so striking that
its simplification is the main goal of electronic modeling. Hartree-Fock 
(HF), den\-sity-functional (DFT) and reduced-density-matrix
functional (RDMFT) theories attempt to achieve this goal by using,
respectively, one Slater determinant, the electron density or the
one-body reduced density matrix as the basic variable.  

The one-body reduced density matrix is 
obtaining by tracing out $N - 1$ particles, and reads, for
a $N$-fermion quantum state $\ket{\Psi}$,
\begin{equation}
  \hat\gamma \equiv N { \rm Tr}_{N-1}[\ket{\Psi}\langle\Psi|].
\end{equation}
In the quantum-chemistry jargon, the \textit{na\-tu\-ral occupation
  numbers} are the eigenvalues (organized in decreasing order $n_1
\geq n_2 \geq \cdots$) and the \textit{natural spin-orbitals} are the
eigenvectors $\{\ket{\varphi_i}\}$ of $\hat\gamma$.  The theoretical
framework of RDMFT is based on a variational principle stating that
the ground-state energy of a fermionic system can be obtained by
minimizing some energy functional on the set of $N$-representable
one-body reduced density matrices \cite{PhysRevB.12.2111,Pernal2016}.
The (ensemble) $N$-representability conditions of $\hat\gamma$ (the
famous Pauli exclusion principle) depend on its eigenvalues only,
reading simply \cite{Col2}:
\begin{equation}
\label{eq:rep}
0 \leq n_i \leq 1 \qquad {\rm and} \qquad  \sum_i n_i = N,
\end{equation}
where $N$ is the number of electrons of the system under 
consideration.  

For pure quantum systems the occupation numbers meet 
also additional requirements with tremendous physical implications 
 \cite{Kly2,Kly3,CS2013,BenavLiQuasi,Mazz14,
 MazzOpen,BenavQuasi2, CSQuasipinning,C7CP01137G,doi:10.1063/1.5010985}. 
 This so-called {\it generalized} Pauli exclusion principle
 provides a (large) set of constraints on the natural 
 occupation numbers. 
 These are much more stringent than the ordinary Pauli principle, and take the form of independent linear 
 inequalities, namely,
\begin{equation}
  \label{eq:gpc}
  \mathcal{D}_j(\vec{n}) \equiv \kappa_j^{0}+\sum^d_{i = 1}\kappa_j^{i}  
  n_i\geq 0 ,
\end{equation}
where $d$ is the dimension of the one-particle Hilbert space.
The coefficients $\kappa_j^{i}$ are integers. 

In RDMFT the $N$-representability conditions 
(\ref{eq:rep}) can be easily taken into account. Yet, 
the exact correlation functional 
is, by and large, not available and therefore the predicted 
RDMFT energy can be either lower or higher than the exact 
ground-state energy. An exception of this is the M\"uller functional 
\cite{MULLER1984446}, for it is believed that it constitutes 
an universal lower bound for quantum mechanics. So far, 
this statement has been rigorously proved only for two-electron 
systems  \cite{Lieb,QUA:QUA23101}.
To write a correlation functional one often starts by engineering an
approximate expression for the two-body reduced density matrix
\begin{equation}
  \hat\Gamma \equiv \tbinom{N}{2} { \rm Tr}_{N-2}        
   [|\Psi\rangle\langle\Psi|] .
\end{equation}
This is normally accomplished by writing $\hat \Gamma$ 
in terms of the ex\-chan\-ge-correlation hole, 
defined by the following relation:
\begin{equation}
  \Gamma(\xx_1,\xx_2) \equiv \tfrac{1}{2} \rho(\xx_1)[\rho(\xx_2) -
    \rho^{\rm hole}_{\rm xc}(\xx_1,\xx_2)].
\end{equation}
We used the customary
compact notation $\xx \equiv (\rr, \varsigma)$.  The electronic
density, the main object in DFT, is of course the diagonal of the 
one-body reduced density matrix. The M\"uller (also called 
Buijse-Baerends) functional describes the
exchange-correlation hole as the square of a \textit{hole amplitude}
\cite{buijse,GPB}, reading  $|\gamma^{1/2}(\xx_1,\xx_2) / 
\sqrt{\rho(\xx_1)}|^2$.  The functional
then reads:
 \begin{align}
   \label{RDMFT}
  \Gamma_{\rm BB}(\xx_1,\xx_2) &=  \frac12
 \rho(\xx_1)\rho(\xx_2) \\ &  \quad -  \frac12\sum_{ij} \sqrt{n_in_j}
  \chi_{ij}(\xx_1)\chi_{ji}(\xx_2),
   \nonumber 
 \end{align}
where 
 $\chi_{ij}(\xx) \equiv \varphi_i(\xx)\varphi_j(\xx)$, assuming from now
 on real natural orbitals.
 Further developments in RDMFT were inspired by this functional.
 Chief among them, the Goe\-de\-cker-Um\-ri\-gar functional is 
 an extension of~\eqref{RDMFT}, 
 excluding self-in\-ter\-actions 
 in the exchange-correlation and the direct Coulomb terms \cite{GU}.
 Another example is the ``power'' functional~\cite{PhysRevA.79.040501}, proposed by Hardy Gross and collaborators, that replaces the square root by a general fractional power.
 Cioslowski and Pernal \cite{CioPer} and Cs\'anyi, Goedecker and 
 Arias \cite{CGA} have all proposed different 
 generalizations based on a distinction between strongly and weakly 
 occupied natural orbitals. 
 A different perspective is given by the study of the 
cumulant part of $\Gamma$ (i.e., $\Gamma - \tfrac12 \gamma 
\wedge \gamma$) under some of its known
representability conditions \cite{eltit, PhysRevLett.119.063002}. 

In all these functionals, the exchange-correlation term of 
the energy functional can be cast into the simple form
\begin{equation}
\label{eq:xc}
\mathcal{E}_{\rm xc}[\gamma] 
= -\frac12 \sum_{ij\varsigma} \int d^3r d^3r' 
f^\varsigma_{ij}(\vec n) 
\frac{\chi_{ij}(\rr,\varsigma)\chi_{ji}(\rr',\varsigma)}{|\rr - \rr'|}.
\end{equation}
Almost all functionals fare quite well in benchmarking tests, yielding errors for 
the correlation at least an order of magnitude~\cite{LM}
smaller than B3LYP~\cite{Stephens1994_11623}, perhaps the most used DFT functional. RDMFT has
also succeeded in predicting more accurate gaps of conventional
semiconductors than semi-local DFT does and has demonstrated
insulating behavior for Mott-type insulators \cite{Sharma,Pernalrev}, another major result stemming from the research group of Hardy Gross.

Unfortunately, most RDMFT functionals were desig\-ned having in mind
singlet ground states. Furthermore, at zero temperature one can argue
that the representability conditions \eqref{eq:rep} are
unsatisfactory, and that different results could be obtained if more
(pure-representability) constraints were imposed \cite{Valone}. This
is especially true in the framework of finite basis sets
\cite{doi:10.1021/acs.jctc.7b00562}.  For this reason, the enforcement
of additional constraints can only improve the total energy
\cite{RDMFT,Eugene}.  Based on recent progress on fer\-mionic exchange
symmetry and, in particular, on the generalization of Pauli exclusion
principle, our aim in this paper is to introduce functionals motivated
by pure representability considerations.

The paper is organized as follows: Section \ref{sec:2} 
discusses the recent solution of the pure 
$N$-representability solution of the one-body reduced 
density matrix, and some of its remarkable physical 
implications. Section \ref{sec:3} presents two RDMFT 
functionals for the Borland-Dennis state, i.e.~the pinned state for three 
fermions in a six dimensional one-particle Hilbert space (say, three 
fermions in six modes). In Section \ref{sec:4} we generalize our 
results for a system with three active (valence) electrons 
and an arbitrary number of modes. In Section 
\ref{sec:5} we test the functionals for Hubbard models
and discuss the numerical quality of the results. 
The paper ends with a conclusion and two appendixes.

\section{Pure representability conditions and stability of the selection rules}
\label{sec:2}
 
For pure systems, the fermionic natural occupation numbers satisfy
sets of generalized Pauli constraints of the form
\eqref{eq:gpc}. Together with the non-increasing ordering of these
numbers, these constraints define a polytope $\mathcal{P}_{N,d}$ in
$\mathbb{R}^d$ for the occupation numbers compatible with pure states
of $N$ fermions in an one-particle Hilbert space of dimension $d$
\cite{Sawicki2011}. The Hartree-Fock point, i.e.~$\ket{\varphi_1
  \cdots\varphi_N}$, lies in one of the vertexes of the polytope.  The
asymptotic properties of such polytopes are actively being
researched~\cite{Tomasz}.  For a recent review of the generalized
Pauli exclusion principle and its growing impact in quantum chemistry
and condensed matter physics we refer to \cite{progress}.

 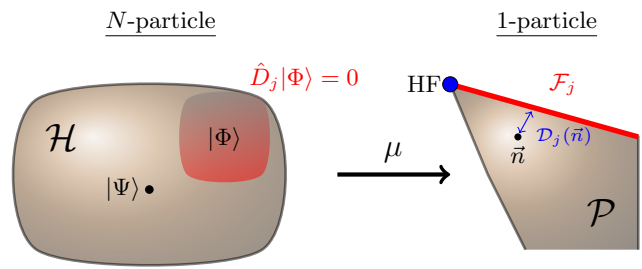
\begin{figure}[b]
        \centering
            \begin{tikzpicture}
              \coordinate (dS) at (1,1.5);
              \shade [ball color=black!10!brown,opacity=0.50,line width=1,draw=black] 
                   plot [smooth cycle] coordinates  
                    { (2,0) (-1,0) (-1,2) (2,2)};
              \draw (-1,1.5) node[right]{\Large $\H$};  
              \draw (-0.2,3) node[right]{\underline{$N$-particle}};  
              \draw (5,3) node[right]{\underline{$1$-particle}};      
              % ploska dS      
              \shade[bottom color=red!10!red,opacity=0.50, shading angle=10](dS) 
                    plot [smooth cycle] coordinates  
                    { (1,1) (1,2) (2,2) (2,1)};
              % text
              \draw [fill=green] (1.5,1.2) node[above, black]
              {$\ket{\Phi}$};
              \draw [fill=black] (0.5,0.8) circle (0.05) node[left,
              black] {$\ket{\Psi}$};
              %\draw (1.2,0.9) -- (1.6,1.6);
              \draw [ultra thick, ->] (3,1) -- (4.5,1);
              \draw [fill=black] (3.4,2.3) circle (0) node[left, black]{\color{red} $\hat{D}_j \ket{\Phi} = 0$};
              \draw (3.5,1.3) node[right] {\large$\mu$};   
              % line - division of object             
             \coordinate (dP) at (2,1.5);
              \shade [ball color=black!10!brown,opacity=0.50,line width=1,draw=black] 
                   plot [tension=2] coordinates  
                    {(5.5,0) (5,1) (4.5,2.2) (7,1.5) (7,0)};         
              \shade [ball color=red!10!red,opacity=4,line width=2,draw=red] 
                   plot [tension=2] coordinates  
                    {(4.5,2.2) (7,1.5)};
               \draw [fill=black] (6.3,2.2) circle (0) node[left, black]{\color{red} $\mathcal{F}_j$};
              \draw [blue, <->] (5.42,1.55) -- (5.58,1.84);
              \draw [fill=black] (6.5,1.5) circle (0) node[left, black]{\color{blue} {\scriptsize $\mathcal{D}_j({\vec n})$}};
              \draw (6.2,0.5) node[right]{\Large $\mathcal{P}$}; 
              \draw [fill=black] (5.4,1.5) circle (0.04) node[below, black]{$\vec{n}$}; 
              \draw [fill=blue] (4.5,2.2) circle (0.1) node[left, black]
              {HF};  
            \end{tikzpicture}
\caption{
Schematic view of the reconstruction of $N$-particle
states based on one-particle information only. $\H$ 
is the fermionic Hilbert space of wave functions of $N$ fermions.
$\mathcal{P}$ is the polytope of pure-representable 
one-particle states. The Hartree-Fock point is represented
as a dot on one of vertexes of the polytope. The 
arrow goes then from $\H$ to 
$\mathcal{P}$ ($\mu: \H \rightarrow \mathcal{P}$). The 
image of the state $\ket{\Psi}$ is then $\mu(\ket{\Psi}) = 
\vec{n}$. The spectra lying on one of the facets $\mathcal{F}_j$
(i.e.~the sets of occupation numbers with 
$\mathcal{D}_{j}(\vec n) = 0$) 
correspond to states satisfying $\hat{D}_{j} \ket{\Phi} = 0$.}
 \label{graf:Reconst}
 \end{figure}
 
The generalized Pauli principle is 
particularly relevant whenever the set of natural occupation 
numbers of a given fer\-mio\-nic state  ``saturates'' a 
generalized Pauli constraint, i.e., the equality holds in 
Eq.~\eqref{eq:gpc}.
 This so-called \textit{pinning} effect is connected with a 
remarkable simplification of the global structure of the 
many-body wave function. In fact, any $N$-fermion state $\ket{\Phi}$, 
with occupation numbers $\vec{n} = (n_1, n_2, \cdots)$, compatible with the 
pinning condition $\mathcal{D}_j(\vec{n}) = 0$, 
belongs to the null eigenspace of 
the operator
\begin{equation}
  \hat{D}_j \equiv \kappa_j^{0}+\sum_{i}\kappa_j^{i}  
  \hat n_i,
\end{equation}
where $\hat n_i$ denotes the number operator of the 
natural orbital $\ket{\varphi_i}$ of $\ket{\Phi}$.
For non-degenerate natural occupation numbers, 
the condition $ \hat{D}_j  \ket{\Phi} = 0$ 
amounts to a simple selection rule for the 
con\-fi\-gu\-rations present in the expansion of the 
quantum state. Indeed, the configuration interaction
 expansion of a \textit{pinned} 
wave function 
\begin{align}
\ket{\Phi} = \sum_{(i_1, \cdots, i_N) \in \mathcal{I}_j} c_{i_1 \cdots i_N} \ket{\varphi_{i_1}\cdots\varphi_{i_N}},
\label{eq:simpli}
\end{align}
where $\ket{\varphi_{i_1}\cdots\varphi_{i_N}}$ denotes 
a normalized Slater 
determinant, is restricted to 
configurations 
belonging to $\mathcal{I}_j$, 
namely, the set of 
determinants fulfilling the 
selection rule
\begin{align}
\hat{D}_j \ket{\varphi_{i_1}
\cdots\varphi_{i_N}} = 0.
\end{align}
For a schematic view of the wa\-ve-function 
reconstruction see Fig.~\ref{graf:Reconst}.
 To give an example of these representability constraints, 
the rank-six approximation (i.e., six modes or six natural 
spin-orbitals) for the three-electron system
(the so-called Bor\-land-Dennis \textit{setting}) is completely 
characterized by four such constrains~\cite{Borl1972}, namely,
the equalities $n_{1} + n_{6} = n_{2} + n_{5} = 
n_{3} + n_{4} = 1$ and the inequality:
\begin{align}
   \mathcal{D}_{\rm BD}(\vec{n}) = 2- n_{1} - n_{2} - n_{4} \geq 0 . 
  \label{eq:BD2}
\end{align}
This latter inequality together with the non-increasing
ordering of the natural occupation numbers defines 
a polytope in $\mathbb{R}^6$, called the 
``Borland-Dennis Paulitope''. The\-se conditions imply 
that, in the natural orbital 
basis, every Slater determinant 
$\ket{\varphi_{i}\varphi_{j}\varphi_{k}}$, 
built up from three natural spin-orbitals, 
showing up in the configuration interaction
expansion of the Borland-Dennis setting, satisfies  
\begin{align}
\ket{\varphi_i\varphi_j\varphi_k} =
(\hat n_{7-s} + \hat n_{s})\ket{\varphi_i\varphi_j\varphi_k}
\end{align}
for $s\in\{1,2,3\}$. 
The Borland-Dennis \textit{state}, in addition, 
fullfils $\mathcal{D}_{\rm BD}(\vec{n}) = 0$,
and therefore according to \eqref{eq:simpli},
reads:
\begin{align}
\ket{\Phi_{\rm BD}} = \alpha  \ket{\varphi_{1}
\varphi_{2}\varphi_{3}} +
\beta \ket{\varphi_{1}\varphi_{4}\varphi_{5}} + 
\gamma  \ket{\varphi_{2}\varphi_{4}\varphi_{6}}.
\label{eq:bd1}
\end{align}

Noticeably, the selection rule for pinned states 
is stable in the sense that being in the vicinity
of the Paulitope boundary ($\mathcal{D}_j(\vec{n}) \approx 0$) 
implies approximately the simplified structure \eqref{eq:simpli}. 
This important result states that, in other words, any many-fermion 
quantum state $\ket{\Psi}$ can be approximated by the 
structural simplified form \eqref{eq:simpli}, corresponding to 
saturation of the generalized Pauli constraint $\mathcal{D}_j$, 
up to an error bounded by the distance of $\vec{n}$ to the 
corresponding polytope facet \cite{SBV}:
\begin{align}
1 - ||\hat P_{j} \Psi||^2 \leq 2  \mathcal{D}_j(\vec{n}).
\end{align}
Here $\hat P_j$ is the projector on the zero-eigenspace of 
$\hat D_j$.

There is a growing corpus of theoretical and 
numerical evidence that indicate that,
for some systems, ground states are quasipinned to one or 
more boundaries of the pertinent polytope 
\cite{BenavQuasi2,CS2015Hubbard,CSHFZPC,CS2016b}.
It is therefore reasonable to assume that such ground states 
have approximately a simplified structure due to pinning.
Inspired by this result, our main aim in this paper is 
to produce systematically functionals for RDMFT 
for quantum systems very close to the boundary of 
the polytope.
 
A word of caution is in order here. To some extent, the interplay
between RDMFT functionals and pure $N$-re\-pre\-sentability is 
not a happy tale. Some time ago, Pernal \cite{PERNAL2013127} discovered that
the so-called PNOF5 functional \cite{doi:10.1063/1.4844075} is
strictly pure $N$-representable. Although this result displayed the
mathematical quality of the functional, it also showed that for this
reason it underestimated seriously dynamic electron correlation. Our
functionals are inspired on the structural simplification
\eqref{eq:simpli} but we do not require that the final result is
representable. We are just orienting our search for a RDMFT functional
in order to capture correctly strong correlation, at least for finite
Hilbert spaces.

\section{Three active electrons}
\label{sec:3}

We consider here the Borland-Dennis wave function 
$\ket{\Phi_{\rm BD}}$ given in Eq.~\eqref{eq:bd1},
which in terms of the natural occupation numbers reads:
\begin{align}
\label{LSs}
\ket{\Phi_{\rm BD}[\vec{n},\vec\varphi]} &= \sqrt{n_{3}} \, \ket{\varphi_{1}
\varphi_{2}\varphi_{3}}  \\ &\quad -
 \sqrt{n_{5}} \,
\ket{\varphi_{1}\varphi_{4}\varphi_{5}} + \sqrt{n_{6}} \, 
\ket{\varphi_{2}\varphi_{4}\varphi_{6}}, \nonumber
\end{align}
with $\vec \varphi \equiv (\varphi_1, \varphi_2, \cdots)$.
By normalization, $n_{3} + n_{5} + n_{6} = 1$.
Remarkably, just like in the famous L\"owdin-Shull functional, 
exact for two-fermion systems~\cite{LS}, the wave function 
\eqref{LSs} ---only exact for a pinned three-electron system 
within the rank-six approximation--- is 
explicitly written in terms of both the natural occupation 
numbers and the natural orbitals. Hence, the Borland-Dennis
state is by itself a functional of these quantities.
Likewise, any sign dilemma that may occur when writing the 
amplitudes of the states
\eqref{LSs} can be circumvented by absorbing the possible 
phases into the spin-orbitals. In addition, only doubly 
excited configurations are permitted. 

Remember that the occupation numbers also satisfy 
the pinning constraints $n_i + n_{7-i} = 1$.
Just for convenience we choose a negative sign in front
of the second Slater determinant, which can be viewed, without
loss of generality, as a rotation of the fifth natural orbital. 
 The two-body reduced-density matrix for the Borland-Dennis
 state can be separated in two terms: the diagonal part
\begin{align}
\hat \Gamma^{(d)}_{\rm BD}[\vec{n},\vec\varphi] =
\sum_{k\in\{3,5,6\}}
n_{k} 
\sum_{i < j \in \mathcal{Z}_k} 
\ket{\varphi_i\varphi_j}\bra{\varphi_i\varphi_j},
\end{align}
where
$\mathcal{Z}_3 = \{1,2,3\}$, $\mathcal{Z}_5 = \{1,4,5\}$ and
$\mathcal{Z}_6 = \{2,4,6\}$, and the non-diagonal one
\begin{align}
\hat \Gamma^{(nd)}_{\rm BD}[\vec{n},\vec\varphi] =& - \sqrt{n_{3} n_{5}} \nonumber
(\ket{\varphi_{2}\varphi_{3}}\bra{\varphi_{4}\varphi_{5}} + h.c.) 
\\ & -  \sqrt{n_{5} n_{6}} (\ket{\varphi_{1}\varphi_{5}}\bra{\varphi_{2}\varphi_{6} } + h.c.)
\nonumber  \\ 
& -  \sqrt{n_{3} n_{6}} (\ket{\varphi_{1}\varphi_{3}}\bra{\varphi_{4}\varphi_{6}} + h.c.) .
\label{eq:nondiagonal}
\end{align}
This latter term is responsible, 
so to speak, for the pure character of $\hat \Gamma_{\rm BD}$. 
In fact, without this term, the Borland-Dennis state would reduce to an incoherent 
 superposition of Slater determinants; therefore, a mixed state. 
 The appearance of the terms 
 \begin{align}
 \varphi_{i}(\xx_1)\varphi_{j}(\xx_1)\varphi_{k}(\xx_2)\varphi_{l}(\xx_2)
 \end{align}
 in \eqref{eq:nondiagonal} is by no means new in the realm of
 RDMFT. 
Recently, Gebauer, Cohen and Car introduced a linear scaling 
functional containing such terms \cite{Gebauer15112016}. 
These new terms lead us to represent the 
exchange-correlation functional \eqref{eq:xc} 
in a more general fashion:
\begin{align} 
\mathcal{E}_{\rm xc}[\vec{n},\vec\varphi] = -\frac12 \sum_{ijkl} \int d^3x d^3x' 
f^{\varsigma\varsigma'}_{ijkl}(\vec{n}) 
\frac{\chi_{ij}(\xx)\chi_{kl}(\xx')}{|\rr - \rr'|}.
\label{newf}
\end{align}
Obviously, $f^{\varsigma\varsigma'}_{ijkl}(\vec{n})
= f^{\varsigma'\varsigma}_{klij}(\vec{n})$. It is worth mentioning
that there is an important 
physical motivation in writing $\mathcal{E}_{\rm xc}$ 
in this way. In fact, the new functional \eqref{newf} is
in principle spin inseparable,  
namely,  $\mathcal{E}_{\rm xc}[\gamma] \neq
\mathcal{E}_{\rm xc}[\gamma^\up] 
+ \mathcal{E}_{\rm xc}[\gamma^\dn]$. It is known that
spin separability is not able to reproduce spin polarizations
 and is therefore physically inexact \cite{PhysRevA.96.062508}.

For the doublet, the two-body reduced density matrix  can be 
regarded as a $4\times 4$ matrix in spin space 
\cite{PhysRevA.87.022118}, but due to particle conservation 
only four terms are different from zero: 
\begin{align}
\hat \Gamma = 
\begin{pmatrix}
 \hat \Gamma^{\up\up} & \hat \Gamma^{\up\dn} \\
 \hat \Gamma^{\dn\up} &  \hat \Gamma^{\dn\dn}  \\
\end{pmatrix}.
\end{align}
For the three-particle system, the 
$\hat \Gamma^{\dn\dn}$ term is zero. By symmetry, $ \Gamma^{\downarrow\uparrow}(\rr_1,\rr_2)
= \Gamma^{\uparrow\downarrow}(\rr_2,\rr_1)$.

\subsection{One frozen electron}
\label{eq:doss}

A natural spin orbital is a direct product of a 
spatial orbital and a spinor $\ket{\varsigma}\otimes
\ket{\phi^\varsigma_i}$. The active space is then
described by two sets of orthonormal spatial 
functions, namely, $\{\phi^\up_i\}$ and $\{\phi^\dn_i\}$. Let us 
first consider one electron frozen in the doublet spin 
configuration of the three-electron system.  Save a 
sign indeterminacy, which in principle cannot be removed, 
the wave function reads: 
\begin{align}
\label{appuno}
\ket{\Psi_2[\vec{n},\vec\phi^\varsigma]} = \sum_{\mu>0} (\pm) 
\sqrt{n_\mu} \ket{{\uparrow}{\phi_{0}^{\uparrow},
\uparrow}{\phi_\mu^\uparrow,\downarrow}\phi_\mu^\downarrow }.
\end{align}
It is easy to show that the corresponding elements of
the two-body density matrix read:
\begin{align}
\hat \Gamma^{\uparrow\uparrow}_{2}[\vec{n},\vec\phi^\varsigma] =  
\sum_{\mu> 0} n_\mu
\ket{\phi_0^\uparrow
\phi_\mu^\uparrow}
\bra{\phi_0^\uparrow
\phi_\mu^\uparrow}
\end{align}
and 
\begin{align} \nonumber
\Gamma_{2}^{\uparrow\downarrow}(\rr_1,\rr_2) &=  
\frac12 \sum_{\mu> 0} n_\mu
\chi^{\uparrow}_{00}(\rr_1)
\chi_{\mu}^{\downarrow}(\rr_2) \\ &+ \frac12 
\sum_{\mu,\nu > 0} (\pm) \sqrt{n_\mu n_\nu} 
\chi^{\uparrow}_{\mu\nu}(\rr_1)
\chi_{\nu\mu}^{\downarrow}(\rr_2) .
\label{eq:five} 
\end{align}
It is instructive to realize that the last term of Eq.~\eqref{eq:five}
is the L\"owdin-Shull functional for two active electrons. The final
expression of the exchange-correlation functional can be written using
similar approximations as Buijse and Baerends used for their
functional \cite{buijse}. The first approximation uses the fact that
the coefficients $\sqrt{n_\mu}$ are usually very small for $\mu > 1$,
allowing us to neglect the product of
two such terms. The second approximation addresses the sign
undeterminancy by choosing a negative sign in front of these small
coefficients and a positive one in front of the first dominant term.
This choice is justified by the positivity of the Coulomb potential
which ensures that the off-diagonal terms ($p > 1$)
\begin{align}
\int \frac{\chi_{1p}(r)\chi_{p1}(r')}{|r-r'|}dr dr'
\end{align}
are positive. By the Rayleigh principle, a lower value of the 
ground-state energy will be reached only if the first sign is negative
\cite{QUA:QUA23101}. 
In this way, the exchange-correlation functional reads
\begin{align}
f^{\uparrow\uparrow}_{2,0\mu\mu0} (\vec{n}) & = 
f^{\uparrow\uparrow}_{2,\mu00\mu} (\vec{n}) =
n_\mu, \quad \;\forall \mu \geq 0, \\
f^{\uparrow\uparrow}_{2,\mu\mu\nu\nu} (\vec{n}) & = 
n_\mu n_\nu, \qquad  \qquad \qquad \forall \mu, \nu > 0,
\nonumber
\end{align}
with $n_0 = 1$.
Furthermore, like in the M\"uller functional,
\begin{align}
f^{\uparrow\downarrow}_{2,\mu\nu\nu\mu} (\vec{n}) = \sqrt{n_\mu n_\nu},
\quad \forall \mu,\nu > 0.
\end{align}
Finally, to correctly cancel the spurious self-interaction
contribution of the Coulomb term
$
f^{\dn\dn}_{2,\mu\mu\nu\nu}
(\vec{n}) = n_\mu n_\nu$.
The remaining terms are zero.

\subsection{Borland-Dennis functional}
\label{eq:BDs}

In this section we write the doublet configuration of the 
Borland-Dennis state \eqref{LSs} in terms of the spatial 
orbitals explicitly:
\begin{align}
\label{eq:BD}
\ket{\Psi_{\rm BD}[\vec{n},\vec\phi^\varsigma]} &= \sqrt{n_1} 
\ket{{\uparrow}{\phi_0^\uparrow,\downarrow} {\phi_1^\downarrow,\uparrow} \phi_1^\uparrow}
 \\ & -  \sqrt{n_2} \ket{{\uparrow}{\phi_0^\uparrow,\downarrow} {\phi_2^\downarrow,  \uparrow}\phi_2^\uparrow} 
 + \sqrt{n_3} \ket{{\uparrow}{\phi_1^\uparrow,\downarrow}
 {\phi_3^\downarrow, \uparrow} \phi_2^\uparrow}.
  \nonumber
\end{align}
The corresponding two-body reduced density matrix is
\begin{align}
\nonumber
\hat \Gamma_{\rm BD}^{\uparrow\uparrow} &=
n_1 
\ket{\phi_{0}^\uparrow
\phi_{1}^\uparrow}
\bra{\phi_{0}^\uparrow
\phi_{1}^\uparrow} 
\\ & \quad+
n_2
\ket{\phi_{0}^\uparrow
\phi_{2}^\uparrow}
\bra{\phi_{0}^\uparrow
\phi_{2}^\uparrow}
+ n_3 
\ket{\phi_{1}^\uparrow
\phi_{2}^\uparrow}
\bra{\phi_{1}^\uparrow
\phi_{2}^\uparrow}.
\label{twoupup}
\end{align}
Notice that the first two terms of \eqref{twoupup} also 
appear in the density matrix for the 
rank-five approximation $\hat \Gamma^{\uparrow\uparrow}_{2}$.
The non-diagonal parts of $\hat \Gamma_{\rm BD}$
are given by
\begin{align}
\nonumber
 \Gamma_{\rm BD}^{\uparrow\downarrow}(\rr_1,\rr_2)  
& = \frac{n_1}2 [\chi^\uparrow_{00}(\rr_1) \nonumber
+ \chi^\uparrow_{11}(\rr_1)] \chi^\downarrow_{11}(\rr_2) \\
& + 
\frac{n_2}2 [\chi^\uparrow_{00}(\rr_1) \nonumber
+ \chi^\uparrow_{22}(\rr_1)] \chi^\downarrow_{22}(\rr_2) \\
& 
+ \frac{n_3}2 [\chi^\uparrow_{11}(\rr_1) \nonumber
  + \chi^\uparrow_{22}(\rr_1)] \chi^\downarrow_{33}(\rr_2)  \\
 & - 
 \Delta^{\uparrow\downarrow}(\rr_1,\rr_2).
\end{align}
The last term contains the following exchange terms:
\begin{align}
\label{twi}
&\Delta^{\uparrow\downarrow}(\rr_1,\rr_2) = \sqrt{n_1n_2} 
\chi^\uparrow_{12}(\rr_1) \chi^\downarrow_{21}(\rr_2)\\ 
&\quad+ \sqrt{n_1n_3} 
\chi^\uparrow_{02}(\rr_1) \chi^\downarrow_{13}(\rr_2)
+\sqrt{n_2n_3} 
\chi^\uparrow_{01}(\rr_1) \chi^\downarrow_{23}(\rr_2).
\nonumber 
\end{align}
It is interesting to note that only the first term appearing in
Eq.~\eqref{twi} appears also in  \eqref{eq:five}. The other
two correspond to the exchange terms of the Borland-Dennis
wave function. The functional then reads (see Appendix \ref{eq:appendic}): 
\begin{align}
\label{equnof}
f^{\uparrow\uparrow}_{{\rm BD},ijji} (\vec n) = \bigg\{
\begin{array}{cl}
n_{i+j}  & \text{if } i \neq j \\
(n_1 + n_{i+2})^2 & \text{if } i = j \in \{0,1\}, \\
\end{array} 
\end{align}
with $i,j \in \{0,1,2\}$. Also, 
$f^{\uparrow\uparrow}_{{\rm BD},1122} = 
f^{\uparrow\uparrow}_{{\rm BD},2211} = n_1 n_3$. 
To correctly cancel the spurious self-interaction 
contribution of the Cou\-lomb term, 
\begin{align}
f^{\dn\dn}_{{\rm BD},\mu\mu\nu\nu} (\vec n) 
= n_\mu n_\nu,
\end{align}
with $\mu,\nu \in \{1,2,3\}$.
Furthermore, 
\begin{align}
f^{\uparrow\downarrow}_{{\rm BD},ii\mu\mu} (\vec n) = 
 \bigg\{
\begin{array}{cl}
-n_{3-i}n_\mu   & \text{if } i + \mu \neq 3 \\
(1 -n_\mu) n_\mu & \text{if }  i + \mu = 3. \\
\end{array} 
\label{eqdosf}
\end{align}
Finally, the exchange term of  $\hat \Gamma^{\up\dn}$
is nothing more than the
 ``twisted'' exchange term 
 $\Delta^{\uparrow\downarrow}(\rr_1,\rr_2)$.
 
\subsection{New functional}

In this paper we seek a good compromise between
the dynamical-correlated functional $\hat \Gamma_2$ and the 
static-correla\-ted functional $\hat \Gamma_{\rm BD}$. 
A linear superposition of these two density matrices would be 
a good starting point.  Although the result would be 
 not representable, it would fulfill the sum rule $\int \Gamma
 (\xx_1,\xx_2; \xx'_1,\xx_2) d\xx_2
 = \gamma(\xx_1,\xx_1')$. 
 For engineering the exchange-correlation functional 
 we proceed, however, in the following way. Notice 
 that the final outcome of the Sections \ref{eq:doss}
 and \ref{eq:BDs} is the proposal of two exchange-coorelation 
 functionals which share one common term:
\begin{align}
\nonumber
\E_2[\vec{n},\vec\phi^\varsigma] &= \mathcal{F}[\vec{n},\vec\phi^\varsigma] + \mathcal{F}_2[\vec{n},\vec\phi^\varsigma] \\
\E_{\rm BD}[\vec{n},\vec\phi^\varsigma] &= \mathcal{F}[\vec{n},\vec\phi^\varsigma] + \mathcal{F}_{\rm BD}[\vec{n},\vec\phi^\varsigma]. 
\end{align}
Typically these functionals undercorrelate, each representing just a 
fraction of the correlation energy. Assume that the functional 
 $\mathcal{F}[\vec{n},\vec\phi^\varsigma] + 
 \mathcal{F}_{2}[\vec{n},\vec\phi^\varsigma] +
 \mathcal{F}_{\rm BD}[\vec{n},\vec\phi^\varsigma]$ overcorrelates. 
 We seek a constant $\alpha$ such that the new functional 
\begin{align}
\label{eq:funcnew}
\E_{\alpha} [\vec{n},\vec\phi^\varsigma]  &=
  \mathcal{F}[\vec{n},\vec\phi^\varsigma]  \\ & \quad + 
\alpha(\mathcal{F}_2[\vec{n},\vec\phi^\varsigma] 
+\mathcal{F}_{\rm BD}[\vec{n},\vec\phi^\varsigma]) 
\nonumber
\end{align}
gives an energy close to the one of the ground state.
We call this the ``$\alpha$-functional''. The terms of the new 
functional can be easily written. For instance, $\forall \mu,\nu > 0$
\begin{align}\nonumber
&f^{\uparrow\downarrow}_{\mu\nu\nu\mu} (\vec{n}) =   \\
&\quad [\alpha + (\delta^1_\mu\delta^2_\nu
+ \delta^1_\nu\delta^2_\mu + \delta^1_\mu\delta^1_\nu
+ \delta^2_\mu\delta^2_\nu) (1-\alpha)]\sqrt{n_\mu n_\nu},
\end{align}
as well as
$f^{\uparrow\downarrow}_{0213} (\vec{n}) = 
\alpha \sqrt{n_1 n_3}$ and 
$f^{\uparrow\downarrow}_{0123} (\vec{n}) = 
\alpha \sqrt{n_2 n_3}$. 
In Appendix
\ref{eq:appendib} we give an argument on how to compute 
$\alpha$.

It is worth mentioning that 
the fully polarized case can be easily tackled for the Borland-Dennis 
state by separating the natural orbitals in two sets, namely
$\{1,2,4\}$, for which we use latin letters $(a,b...)$,
and $\{3,5,6\}$, for which we use greek letters
$(\mu, \nu,...)$. Similar expressions to \eqref{equnof}
and \eqref{eqdosf} can be easiiy obtained.

\section{Generalizations}
\label{sec:4}

\subsection{Higher-rank representability conditions}

In rank seven there are four generalized Pauli 
constraints 
\begin{align}
\mathcal{D}_1(\vec n)  = 2 - n_2 - n_3 - n_4 - n_5 \geq 0,
\nonumber \\ 
\mathcal{D}_2 (\vec n)  = 2 - n_1 - n_3 - n_4 - n_6 \geq 0,
\nonumber \\ 
\mathcal{D}_3(\vec n)  = 2 - n_1 - n_2 - n_4 - n_7 \geq 0,
\nonumber \\ 
\mathcal{D}_4(\vec n)  = 2 - n_1 - n_2 - n_5 - n_6 \geq 0.
\end{align}
The saturation of the four constraints ($
\mathcal{D}_j(\vec n) = 0$) implies
 the saturation of the lower-rank Borland-Dennis
relation $\mathcal{D}_{\rm BD}(\vec n)= 0$. This can be 
explained in the following way: for the settings of 
$N$ fermions in a $d$-dimensional one-particle Hilbert 
space and $N$ fermions in a $d'$-dimensional one-particle Hilbert 
space, such that $d < d'$, the corresponding polytopes satisfy:  
$\mathcal{P}_{N,d} = {\mathcal{P}_{N,d'}}|
_{n_{d+1} = \cdots = n_{d'} = 0}$.
It means that, intersected with the hyperplane 
given by $n_{d+1} = \cdots = n_{d'} = 0$, the 
polytope $\mathcal{P}_{N,d'}$ coincides with 
$\mathcal{P}_{N,d}$. Therefore
$\mathcal{P}_{3,6} = {\mathcal{P}_{3,7}}|
_{n_{7} = 0}$. In this case $\mathcal{D}_1(\vec n) = 
\mathcal{D}_2(\vec n) = \mathcal{D}_4(\vec n) = 0$
implies $n_7 = 0$ and therefore $\mathcal{D}_3(\vec n) =
\mathcal{D}_{\rm BD}(\vec n) = 0$.

If one generalized constraint is not fixed, one Slater
determinant is added to the Borland-Dennis wave function. 
The rule to add such a determinant is simple:
the natural orbitals in the configuration 
correspond to the occupation numbers not appearing 
in the unsaturated generalized Pauli constraint. 
For instance, if $\mathcal{D}_4(\vec n) \neq 0$ the new 
Slater determinant is 
$\ket{\varphi_{3}\varphi_{4}\varphi_{7}}$.
We do not consider here the unsaturation of the 
third constraint, because we would end up in 
a six-rank wave function. Let us consider 
that $ \mathcal{D}_2(\vec n) = 
\mathcal{D}_3(\vec n) = \mathcal{D}_4(\vec n) = 0$ and
$\mathcal{D}_1(\vec n) \neq 0$, the structure of the 
wave function then reads:
 \begin{align}
\ket{\Phi_{\rm 7}[\vec{n},\vec \varphi]} &= \sqrt{n_{3}} \, \ket{\varphi_{1}
\varphi_{2}\varphi_{3}} \nonumber +
 \sqrt{n_{5}} \,
\ket{\varphi_{1}\varphi_{4}\varphi_{5}} 
\\ &\qquad + \sqrt{\lambda} \, \ket{\varphi_{2}\varphi_{4}\varphi_{6}}
 + \sqrt{n_7}\, \ket{\varphi_{1}\varphi_{6}\varphi_{7}},
\end{align}
where 
$\lambda = 1- n_3 - n_5 - n_7$. This shows that 
$\ket{\Phi_{\rm 7}}$, as for the Borland-Dennis 
state, can be written in terms of the natural orbitals and the 
natural occupation numbers.

\subsection{Frozen electrons}

In quantum chemistry it is customary to separate the
one-particle Hilbert space in core (fully occupied), active
(partially occupied) and virtual (empty) spin-orbitals.  The core
spin-orbitals are pinned (completely populated) and are not treated as
correlated.  For the case of $r$ core (and consequently $d-r$ active
orbitals) the Hilbert space is isomorphic to the wedge product between
a Hilbert space of $r$ electrons in a $r$-dimensional one-particle
Hilbert space and a Hilbert space of $d-r$ electrons in a
$d-r$-dimensional one-particle Hilbert space Hence, the wave function
can be written as $ \ket{\Psi_r} =
\ket{\varphi_1\cdots\varphi_r} \wedge \ket{\Psi^{\rm active}}$, where
$\ket{\Psi^{\rm active}}$ is the part of the wave function belonging
to the space of fractional occupied natural orbitals.  While the first
$r$ natural occupation numbers are saturated to $1$, the remaining
$d-r$ natural occupation numbers $(n_{r +1}, \cdots, n_d)$ satisfy a
set of generalized Pauli constraints and lie therefore inside the
polytope $\mathcal{P}_{N-r,d-r}$ \cite{SBV}.  For the ``active''
Borland-Dennis state we can apply the same considerations discussed in
the last section, namely: if the corresponding constraint
\eqref{eq:BD2} is saturated, the wave function fulfills
$(\hat{n}_{r+1} + \hat{n}_{r+2} + \hat{n}_{r+4})\ket{\Psi} = 2
\ket{\Psi}$, and the set of possible Slater determinants reduces to
just three.  Following our preceding reasoning, the corresponding
two-body density matrix reads
\begin{align}
\hat\Gamma_{r+{\rm BD}} =  \nonumber \sum^r_{i=1}  \sum^{r+6}_{j=1} n_j
\ket{\varphi_{i}\varphi_{j}}\bra{\varphi_{i}\varphi_{j}}
+ \hat\Gamma_{\rm BD},
\end{align}
where we use the fact that $n_1 = \cdots = n_r = 1$.

\section{Numerical investigations}
\label{sec:5}

To illustrate our results we use a simple but non-trivial fermionic
system, namely the few-site Hubbard model. This model is 
capable of exhibiting both weak and strong (static) correlation. 
Besides its importance for solid-state physics, the Hubbard 
model is one of the paradigmatic instances used to simplify 
the description of strongly correlated quantum many-body 
systems and to test RDMFT functionals \cite{PhysRevB.93.085141}.
The Hamiltonian (in second quantization) of the 
one-dimensional $r$-site Hubbard model reads:
\begin{align}
  \label{eq:Hamilt}
\hat H = -\frac{t}2 \sum_{i,\varsigma}
(c^\dagger_{i\varsigma} c_{(i+1)\varsigma}
+ h.c. )
+ 2 U \sum_{i} \hat n_{i\uparrow} \hat n_{i \downarrow},
\end{align}
$i \in \{1,2,\cdots,r\}$.
The operators $c^\dagger_{i\varsigma}$ and $c_{i\varsigma}$ are the fermionic
creation and annihilation operators for a particle on the site $i$
with spin $\varsigma$ and $\hat n_{i\varsigma}$ is the 
particle-number operator. The first term in Eq.~\eqref{eq:Hamilt} 
describes the hopping between two neighboring sites while 
the second represents the on-site interaction.

\begin{figure}[t]
\centering
\includegraphics[width=6cm]{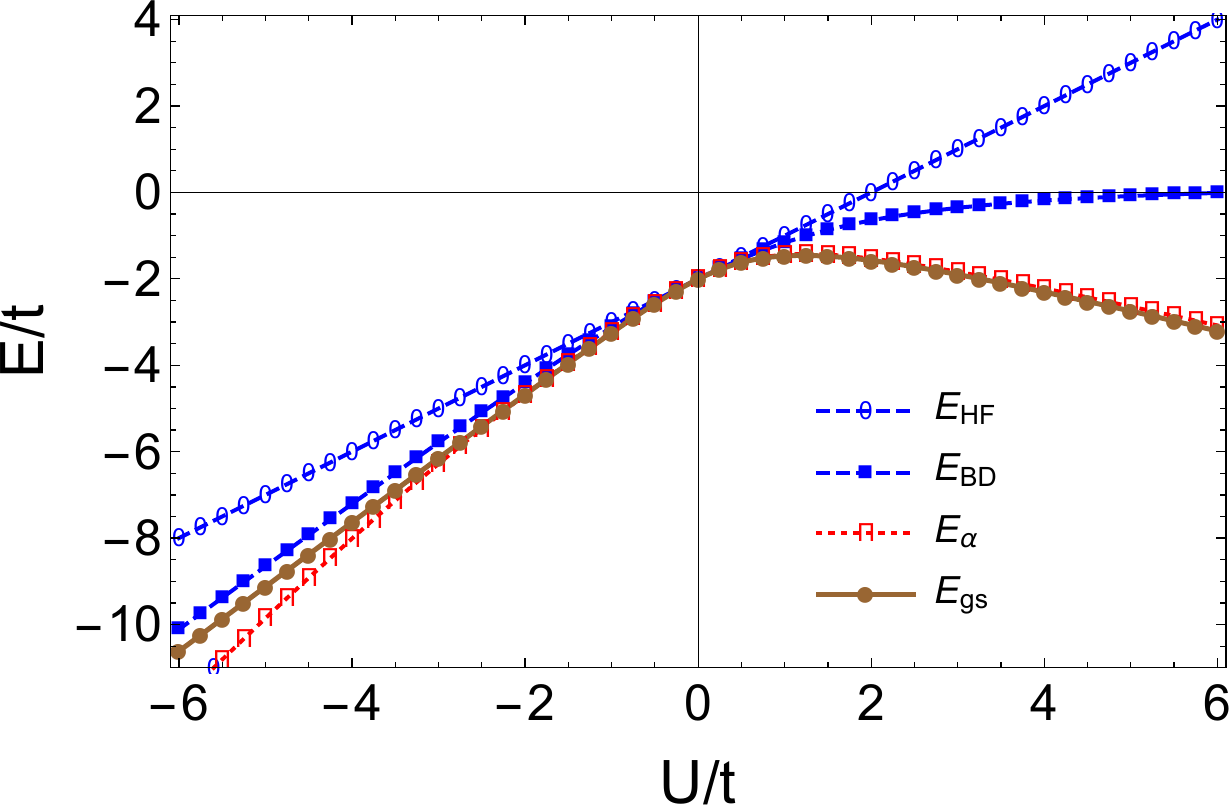}
\includegraphics[width=6cm]{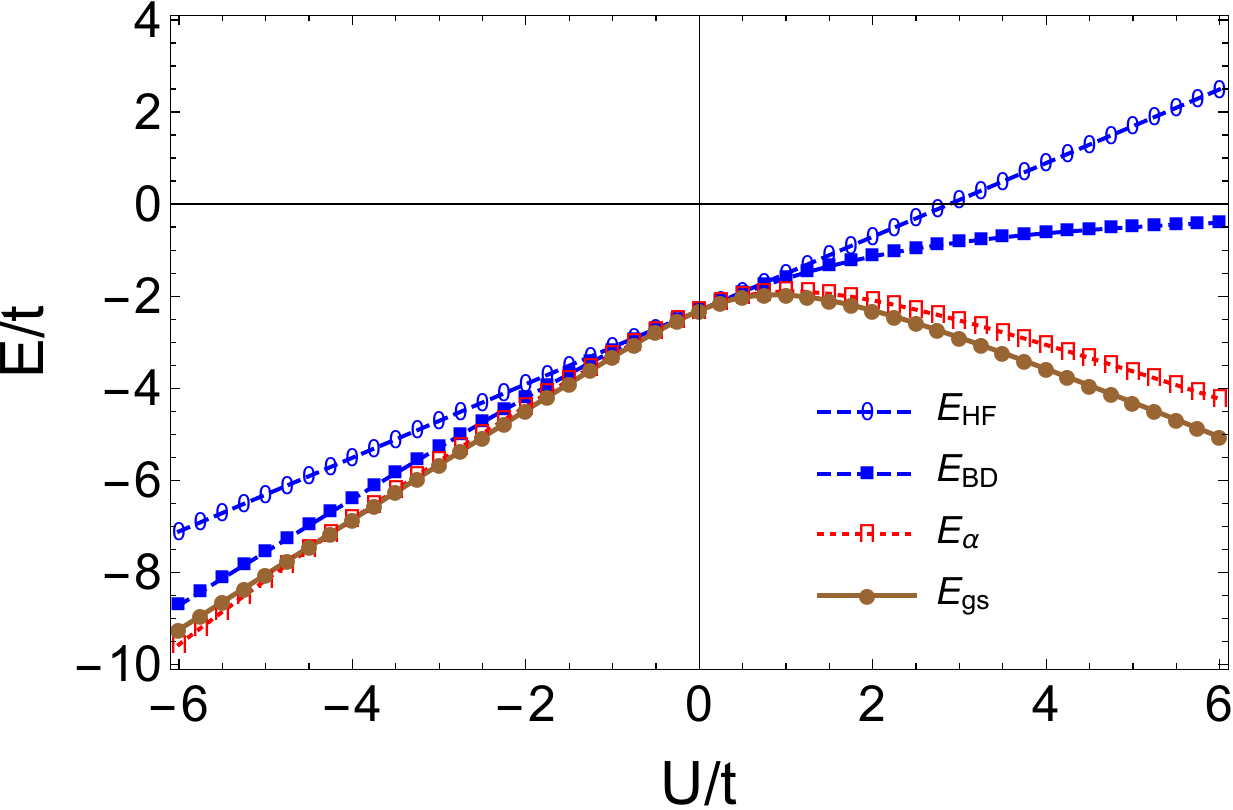}
\caption{For the three-electron Hubbard model we present the energy 
predicted for the Borland-Dennis and the $\alpha$-functionals. 
The upper and bottom panels correspond to the four- and the 
five-site Hubbard model, respectively. 
 The Hartree-Fock and ground-state 
energies are also shown (see text).}
\label{fig:Hubard}
\end{figure}

The one-body reduced density matrix
inherits the symmetries of the corresponding wave function. 
For the Hubbard model with periodic boundary conditions, 
the lattice translation and the $z$-component of the spin 
define such symmetries. 
The spatial part of the natural orbitals is nothing 
more than a Bloch state that satisfies
\begin{align}
\hat T_1 \ket{q} = e^{i\varphi q}\ket{q},
\end{align} 
where $\varphi = 2\pi/r$ and $\hat T_1$ is the one-particle 
translation operator. Using the basis of natural orbitals, the 
energy can be  computed analytically for the rank five 
approximation for the  three-electron Hubbard model
\cite{CS2015Hubbard,newpaper}.  For $r = 4$ 
the result turns out to only depend on the first occupation 
number
\begin{equation}
  E_{r_5}(t,U) = -
  \frac12 ( t - 2 U + \sqrt{9 t^2 + U^2})
\end{equation}
This result is important because it allows us to compute the 
value of $\alpha$. Indeed, for $r=4$ in Eq.~\eqref{foralpha} 
we approximate $\mathcal{G}[\vec{n}_*,\vec\varphi_*] - 
 \mathcal{F}[\vec{n}_*,\vec\varphi_*] = E_{r_5}(t,U)$.
We computed the best $\alpha$ for 4- and 5-site three-fermion
Hubbard model by averaging Eq.~\eqref{foralpha} for $U/t \in [-6,6]$.
 This value turns out to be $\approx 0.72$, which we used from now on.

\begin{figure}[t]
\centering
\includegraphics[width=6cm]{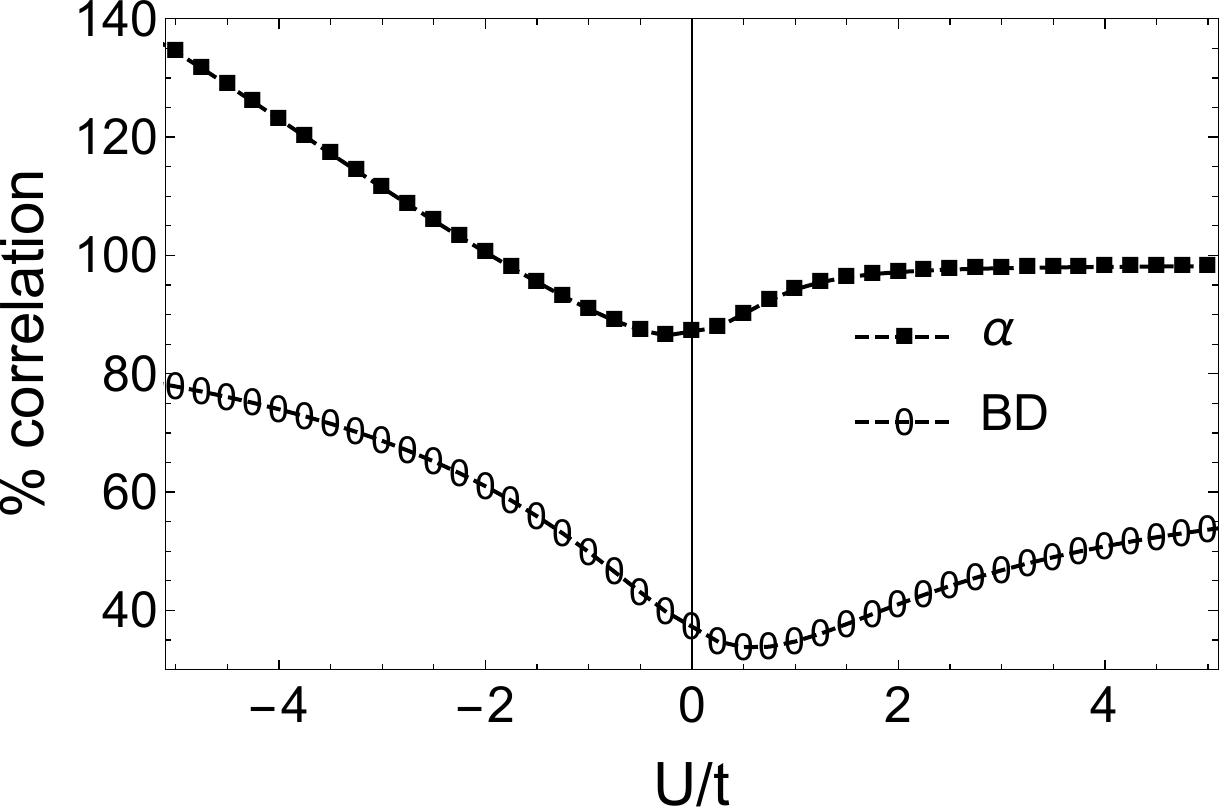}
\includegraphics[width=6cm]{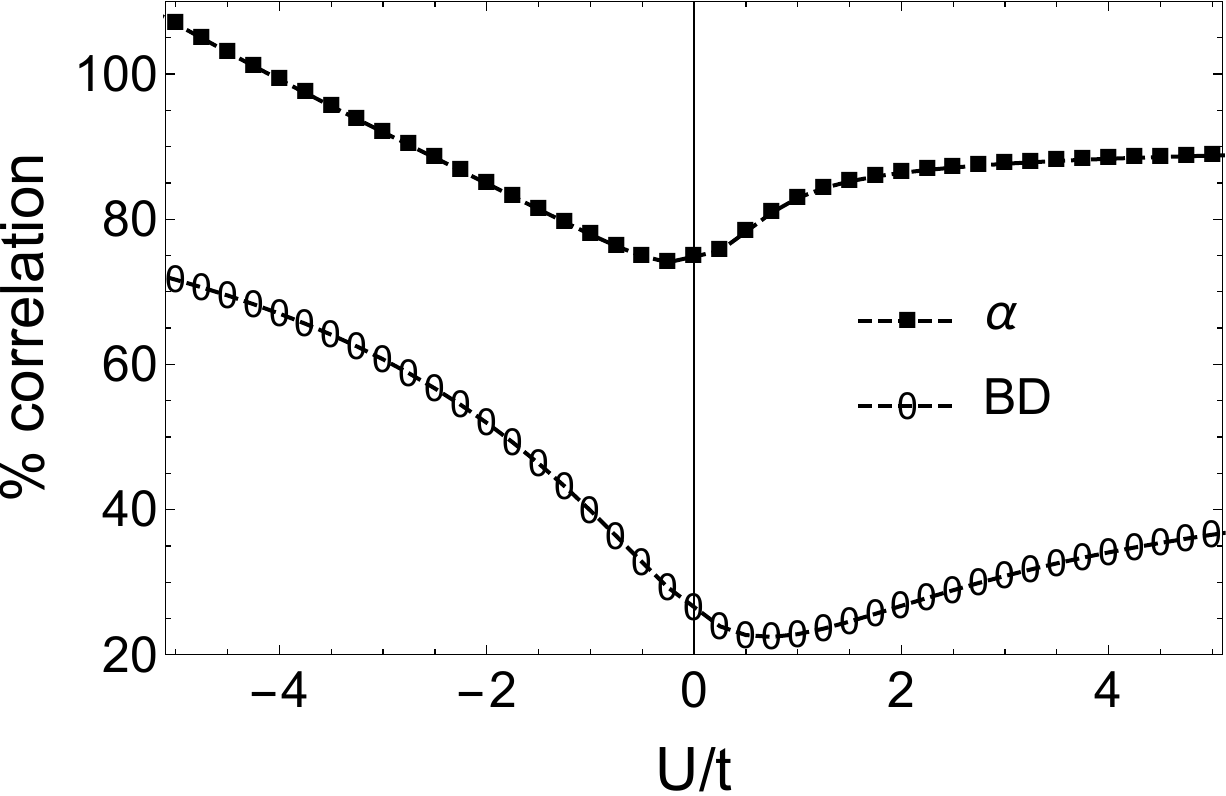}
\caption{For the three-electron Hubbard model we present 
the percentage of the correlation energy recovered by 
the Borland-Dennis and the $\alpha$-functionals.
The percentage is plotted as a function of the relative 
coupling.}
\label{fig:Hubarddos}
\end{figure}

In Figure \ref{fig:Hubard} we present the performance 
of the two functionals introduced in this paper (Borland-Dennis 
and the $\alpha$-functional and compare them with the ground-state 
and Hartree-Fock energies for the four- and five-site Hubbard model. 
For the three-electron problem in the three-site Hubbard model,
the Borland Dennis functional is exact.  It is
worth mentioning the striking performance of the Borland-Dennis 
functional for negative values of the relative strength. 
With our new functional
we recovered almost the full correlation energy. 
Although it slightly overcorrelates for large negative 
values of the relative strength, its performance
is still remarkable (see Fig.~\ref{fig:Hubarddos}). Part of the reason 
for the overcorrelation for negative values is the very low value
of the correlation energy for this sector of interaction.

\section{Conclusions}

Inspired by the recent solution of the pure representability problem
for the one-body reduced density matrix, we proposed a new way of
producing RDMFT functionals. Our approach is based on the specific
simplified structure exhibited by the many-body wave functions whose
occupation numbers are pinned to one or more boundaries of the
polytope of physical states. Some of the states reconstructed in this
way depend explicitly on the natural occupation numbers as well as 
the natural orbitals yielding explicit expressions for the two-body
reduced density matrix. Although this kind of expressions can only be
written for Hilbert spaces of low dimensionality it is possible to
combine this information with other known functionals to derive a
general purpose formula.

Our method leads to a hierarchy of exchange 
correlation functionals, depending to the degree of correlation 
one wishes to reach. In our example, we produced a 
crossbreed functional by combining a dynamical-correlated and 
a static-correlated functional. The results are remarkable even
in the highly correlated regime.  Last but not least, the 
Borland-Dennis functional contains twisted 
exchange-correlation terms of the type $\chi^\varsigma_{kl}(\rr_1) 
\chi^{\varrho}_{ij}(\rr_2)$. These terms, not new but anyhow not common in the realm of RDMFT, are crucial, for example, for 
capturing correctly the time evolution of the system \cite{coming}.

\section*{Acknowledgement}

We thank Nektarios Lathiotakis for helpful discussions.
We acknowledge financial support from the DFG through Projects 
No. SFB-762 and No. MA 6787/1-1 (M.A.L.M.).

\appendix

\section{Borland-Dennis functional}
\label{eq:appendic}

 The one-body reduced density matrix corresponding to the 
 Borland-Dennis state \eqref{eq:BD} is given by the 
 following expressions:
 \begin{align}
\hat \gamma^\up_{\rm BD} &= (n_1 + n_2)
\ket{\phi_0^\uparrow}
 \bra{\phi_0^\uparrow}  \\ & \qquad + (n_1 + n_3)
\ket{\phi_1^\uparrow}
 \bra{\phi_1^\uparrow} + (n_2 + n_3)
\ket{\phi_2^\uparrow} \nonumber
 \bra{\phi_2^\uparrow}
  \end{align}
  and $\hat \gamma^\dn_{\rm BD}  = \sum_i n_i
\ket{\phi_i^\dn}  \bra{\phi_i^\dn} $.
Notice that
 \begin{align}
& \rho^\up_{\rm BD}(\rr_1)  \rho^\up_{\rm BD}(\rr_2)  =   \nonumber
(n_1 + n_2)^2 \chi^\up_{00}(\rr_1)\chi^\up_{00}(\rr_2)\\ 
& \;+
 (n_1 + n_3)^2 \chi^\up_{11}(\rr_1)\chi^\up_{11}(\rr_2) \nonumber
 +  (n_2 + n_3)^2 \chi^\up_{22}(\rr_1)\chi^\up_{22}(\rr_2)  \\ &\;+
 (n_1 + n_2n_3) [\chi^\up_{00}(\rr_1)\chi^\up_{11}(\rr_2)
  \nonumber
 + \chi^\up_{11}(\rr_1)\chi^\up_{00}(\rr_2)]  \\ &\;+
 (n_2 + n_2n_3) [\chi^\up_{00}(\rr_1)\chi^\up_{22}(\rr_2) \nonumber
 + \chi^\up_{22}(\rr_1)\chi^\up_{00}(\rr_2)] \\ &\;+
 (n_3 + n_1n_3) [\chi^\up_{11}(\rr_1)\chi^\up_{22}(\rr_2)
 + \chi^\up_{22}(\rr_1)\chi^\up_{11}(\rr_2)],
  \nonumber
  \end{align}
where we used the normalization condition $n_1 + n_2 + n_3 = 1$.
Self interaction should be canceled, so that 
 \begin{align*}
 &f^{\up\up}_{{\rm BD},0000} = (n_1 + n_2)^2, 
  f^{\up\up}_{{\rm BD},1111} = (n_1 + n_3)^2, \\
 & \qquad  \qquad  f^{\up\up}_{{\rm BD},2222} = (n_2 + n_3)^2.
  \end{align*}
 Notice that, by taking $n_2n_3 \approx 0$, almost all the prefactors
 of the remaining terms of $ \rho^\up_{\rm BD}(\rr_1)  \rho^\up_{\rm BD}(\rr_2)$ show up in \eqref{twoupup} so that the only important 
 term to be added  is 
  \begin{align}
   f^{\up\up}_{{\rm BD},1122} =  f^{\up\up}_{{\rm BD},2211} = n_1 n_3.
   \end{align}
  Furthermore, 
   \begin{align}
 \rho^\dn_{\rm BD}(\rr_1)  \rho^\dn_{\rm BD}(\rr_2)  =  
\sum_{ij} n_in_j \chi^\dn_{ii}(\rr_1)\chi^\dn_{jj}(\rr_2).
  \end{align}
  and therefore 
   $f^{\dn\dn}_{{\rm BD},iijj} =  -n_i n_j$. Finally, 
$\rho^\up_{\rm BD}(\rr_1)  \rho^\dn_{\rm BD}(\rr_2) $
contains terms like
$$
(n_1 + n_2)n_1 \chi^\up_{00}(\rr_1)\chi^\dn_{11}(\rr_2)
= (1  -n_3)n_1 \chi^\up_{00}(\rr_1)\chi^\dn_{11}(\rr_2)
 $$
 A quick inspection
  teaches  us that $f^{\up\dn}_{{\rm BD},ii\mu\mu} = -n_{3-i} n_\mu$, provided that
  $i + \mu \neq 3$. Whenever $i + \mu =3$, 
  \begin{align}
  f^{\up\dn}_{{\rm BD},ii\mu\mu} = (1-n_{\mu}) n_\mu.
  \end{align}
  
 \section{Calculation of $\alpha$}
\label{eq:appendib}

In this Appendix we provide an argument for choosing
$\alpha$. Remember that, for the best value of $\alpha$, 
the minimization of the functional \eqref{eq:funcnew} should 
be approximatively the ground-state energy. Therefore, 
the functional evaluated on the minimizers  ($\vec{n}_*$ and 
$\vec{\varphi}_*$) satisfies
 \begin{align}
 \mathcal{G}[\vec{n}_*,\vec\varphi_*] -  \E_{\rm new} [\vec{n}_*,\vec\varphi_*]  = E_{\rm gs},
 \end{align}
 where 
  \begin{align*}
  \mathcal{G}[\vec{n}_*,\vec\varphi_*] &= 
  \int \delta(\xx-\xx')\hat h \gamma[\vec{n}_*,\varphi_*](\xx,\xx')  d\xx \\ &+ 
 \frac12\int \frac{\gamma[\vec{n}_*,\vec\varphi_*](\xx_1, \xx_1)\gamma[\vec{n}_*,\vec\varphi_*](\xx_2, \xx_2)}{|\rr_1 - \rr_2|} d\xx_1 d\xx_2.
 \end{align*}
 Therefore,
\begin{align}
\alpha =  \frac{\mathcal{G}[\vec{n}_*,\vec\varphi_*] - 
 \mathcal{F}[\vec{n}_*,\vec\varphi_*] - E_{\rm gs}}
 {\mathcal{F}_{2}[\vec{n}_*,\vec\varphi_*]
+\mathcal{F}_{\rm BD}[\vec{n}_*,\vec\varphi_*]}.
\label{foralpha}
\end{align}

\bibliography{Ariadna}

\end{document}